# Quantum Parametric Amplification and Non-Classical Correlations due to 45 nm nMOS Circuitry Effect

Ahmad Salmanogli[1] and Amine Bermak[2]

[2]College of Science and Engineering, Hamad Bin Khalifa University, Doha, Qatar

**Abstract**
This study unveils a groundbreaking exploration of using semiconductor technology in quantum circuitry. Leveraging the unique operability of 45 nm CMOS technology at deep cryogenic temperatures (~ 300 mK), a novel quantum electronic circuit is meticulously designed. Through the intricate coupling of two matching circuits via a 45 nm nMOS transistor, operating as an open quantum system, the circuit's quantum Hamiltonian and the related Heisenberg-Langevin equation are derived, setting the stage for a comprehensive quantum analysis. Central to this investigation are three pivotal coefficients derived, which are the coupling between the coupled oscillator's charge and flux operators through the internal circuit of the transistor. These coefficients emerge as critical determinants, shaping both the circuit's potential as a parametric amplifier and its impact on quantum properties. The study unfolds a delicate interplay between these coefficients, showcasing their profound influence on quantum discord and the gain of the parametric amplifier. Consequently, the assimilation of 45 nm CMOS technology with quantum circuitry makes it possible to potentially bridge the technological gap in quantum computing applications, where the parametric amplifier is a necessary and critical device. The designed novel quantum device serves not only as a quantum parametric amplifier to amplify quantum signals but also enhances the inherent quantum properties of the signals such as non-classicality. Therefore, by carefully engineering the design parameters, one can create an effective parametric amplifier that simultaneously improves the quantum characteristics of the signals. The more interesting result is that if such a theory becomes applicable, the circuit implemented in the deep-cryogenic temperature can be easily compatible with the next step of circuitry while keeping the same electronic features compatibility with the quantum processor.

## Introduction

In the realm of quantum technologies, the need for efficient signal amplification holds paramount importance. Parametric amplifiers, leveraging quantum phenomena to selectively enhance specific frequency components, have emerged as indispensable tools in quantum computing, quantum communication, and quantum sensing applications [1-8]. In essence, parametric amplifiers are a class of devices that hold a special place in the realm of quantum computing. These amplifiers stand out for their unique ability to boost the quantum properties of signals without introducing excessive noise, which is a crucial attribute in the delicate world of quantum information processing [1, 4, 5, 8]. At their core, parametric amplifiers exploit the principles of quantum mechanics to selectively amplify specific quantum states while leaving others untouched. This selective amplification is in stark contrast to classical amplifiers, which amplify all incoming signals uniformly, including noise. In quantum computing, where the preservation of delicate quantum states is paramount, this distinction is invaluable. Parametric amplifiers are designed to operate in a way that they can discriminate between different quantum states within a signal. They achieve this by exploiting the unique properties of quantum systems, such as the ability to be in multiple states simultaneously (superposition) and the sensitivity of quantum systems to specific parameters. The application of parametric amplifiers in quantum technology is multifaceted: 1) In quantum computing, the state of a qubit (quantum bit) needs to be read out with precision. However, quantum states are inherently fragile and often yield weak signals. Parametric amplifiers step in to enhance these signals, enabling accurate and reliable qubit readouts [9]. 2) Quantum communication protocols, such as quantum key distribution, rely on the transmission of quantum signals. Parametric amplifiers bolster the range and fidelity of these signals, enabling secure long-distance quantum communication [10]. 3) In quantum sensing applications, where ultra-sensitive detectors are required, parametric amplifiers serve as invaluable tools. They can amplify the faint signals from quantum sensors, enhancing their precision and sensitivity [11]. 4) Quantum error correction codes are a cornerstone of fault-tolerant quantum computing. Parametric amplifiers aid in error correction by amplifying the signals used to detect and correct errors in quantum codes [12-13]. 5) Parametric amplifiers play a pivotal role in quantum metrology, where precision measurements of physical quantities are paramount. They enable the detection of weak quantum signals with high fidelity, enhancing measurement accuracy [14]. In essence, the parametric amplifier emerges as an indispensable component in the quantum computing toolbox, empowering researchers and engineers to navigate the intricate landscape of quantum states and information. Their unique ability to selectively amplify quantum properties while mitigating noise.

In current quantum systems, high fidelity readout is achieved by using a combination of parametric amplifiers operating at 15 mK and High-Electron-Mobility Transistors (HEMT) LNAs operating at 4.2 K [15-22]. The received and amplified RF signal is then sent to the 300 K RF receiver for further processing. This technology partitioning has the advantage of providing excellent RF noise performance but offers limited opportunities for further integration and poses significant packaging challenges. To reduce packaging costs and increase the level of integration it would be advantageous to integrate the LNA in the CMOS (Complementary Metal-Oxide-Semiconductor). This is the main reason that this article tries to make the parametric amplifier using CMOS technology [23-26]. For this reason, this study delves into the detailed examination and analysis of the two oscillators coupled to nMOS (N-type Metal-Oxide-Semiconductor) transistor to operate as a parametric amplifier, introducing a novel element—a 45 nm nMOS transistor as a pivotal component.

The primary reason behind the adoption of advanced 45nm CMOS technology stems from its exceptional compatibility with quantum processors [24-25]. This seamless integration not only facilitates the creation of a cohesive quantum processing ecosystem but also eliminates the necessity for supplementary technologies such as HEMTs. Furthermore, the ability of CMOS to operate effectively at deep cryogenic temperatures enhances its utility for quantum computing applications, aligning seamlessly with the demands of quantum processing and amplification [23-26].

In this work, by harnessing the interaction coefficients and the intricate internal circuitry of the 45 nm NMOS transistor [27-28], we envision a pathway to designing advanced parametric amplifiers for quantum technologies. This study holds promise in shaping the trajectory of quantum signal amplification and its integration into the broader landscape of quantum computing and sensing. Finally, the confluence of 45 nm CMOS technology and quantum circuit design has yielded a transformative approach to enhancing quantum properties. This work presents not only a theoretical foundation but also a tangible path toward designing innovative quantum devices that flourish under deep cryogenic temperatures. The synergy between cryogenic compatibility, circuit design, and quantum theory holds the potential to revolutionize quantum signal processing, catalyzing the next wave of quantum technological breakthroughs.

## Material and methods

The intricate interplay between the internal circuitry of the transistor and the coupled oscillators introduces a layer of complexity that enables the realization of novel quantum phenomena. The theoretical foundation of our study is grounded in the principles of quantum mechanics. The quantum Hamiltonian governing the coupled oscillators system is derived, capturing the essence of their interaction. The Heisenberg-Langevin equation becomes a vital tool in unveiling the system's dynamics in the quantum regime [29-31]. The theoretical analysis combined with computational techniques, provides insights into the feasibility of designing a parametric amplifier within this unique context. Leveraging the Qutip toolbox in Python [32], we solve the Master equation [29] to calculate critical quantum quantities, including quantum discord, classical discord, quantum purity, and the Symplectic eigenvalue [33-35]. These quantities serve as benchmarks for understanding non-classical correlations and entanglement within the system. Complementary to this, the calculated scattering matrix forms the basis for gain simulations using the input-output formula [29]. This approach allows us to explore how gain enhancement emerges at distinct frequencies, resonating with the oscillators' intrinsic dynamics. Central to this exploration are two critical coefficients: $g_{q1q2}$ and $g_{q2p2}$, the coupling between charge and flux operators through the nMOS internal circuit. These coefficients are unveiled as the keystones to manipulating the system's quantum properties and, in addition, to fostering parametric amplification. The study systematically unveils the delicate interplay between these coefficients and their influence on the system's quantum properties, as well as their impact on the performance of the designed parametric amplifier. By carefully controlling and engineering these coefficients and leveraging the properties of 45 nm CMOS technology, the designed device stands to amplify quantum properties at deep cryogenic temperatures. This breakthrough carries profound implications, potentially obviating the need for traditional signal amplification techniques after quantum processing.

*System definition and study approach*

This study presents the design and analysis of a quantum electronic circuit aimed at investigating the potential of utilizing a 45 nm nMOS transistor as a parametric amplifier. The circuit is constructed by coupling two matching circuits (input and output) via the nMOS transistor, with one matching network connected to the transistor's gate serving as the first oscillator and the other connected to the drain representing the second oscillator (Fig. 1a). The bias circuit is ignored for simplicity. The transistor's source is grounded, and the entire system as an open quantum system (interacted with the reservoir with coupling coefficient $\kappa_{loss} = \kappa_1 + \kappa_2$, where $\kappa_1$ and $\kappa_2$ are the first and second oscillator decay rate, respectively) is analyzed within the framework of quantum theory. For a complete analysis, the internal equivalent circuit of the nMOS transistor is applied [27, 28]. This analysis includes the theoretical derivation of the quantum Hamiltonian and Heisenberg-Langevin equation. In addition, in Fig. 1b and Fig. 1c, $C_{gs}$ changing as the function of $V_{ds}$, and $V_{gs}$ at different corners of operations are illustrated. This point and the related figures will be discussed in the next section.

The primary objective is to explore the viability of employing the 45 nm nMOS transistor as a parametric amplifier, particularly when operated at deep-cryogenic temperatures. By leveraging the internal circuitry of the transistor, the study investigates the existence of coefficients that allow the circuit to function as a parametric amplifier. The relevance of these coefficients to the transistor's internal architecture highlights the significance of controlling thermal noise and designing suitable input and output matching networks. Such manipulation of the circuit's parameters offers the prospect of utilizing the designed system as an effective parametric amplifier. Furthermore, the study delves into the theoretical framework by solving the Master equation, enabling the calculation of critical quantities such as the photon numbers of the oscillators and the exchange of photons facilitated by the transistor's circuitry. Notably, the analysis extends to the exploration of quantum discord, a quantum property that quantifies non-classical correlations. The results underscore the strong influence of coefficients impacting the parametric amplifier's behavior on the quantum discord values. Therefore, by intricately manipulating interaction coefficients and internal circuitry, the potential of enhanced quantum properties and quantum processing is exemplified. The study's insights underscore the viability of utilizing the nMOS transistor as a versatile component within the burgeoning field of quantum technologies.

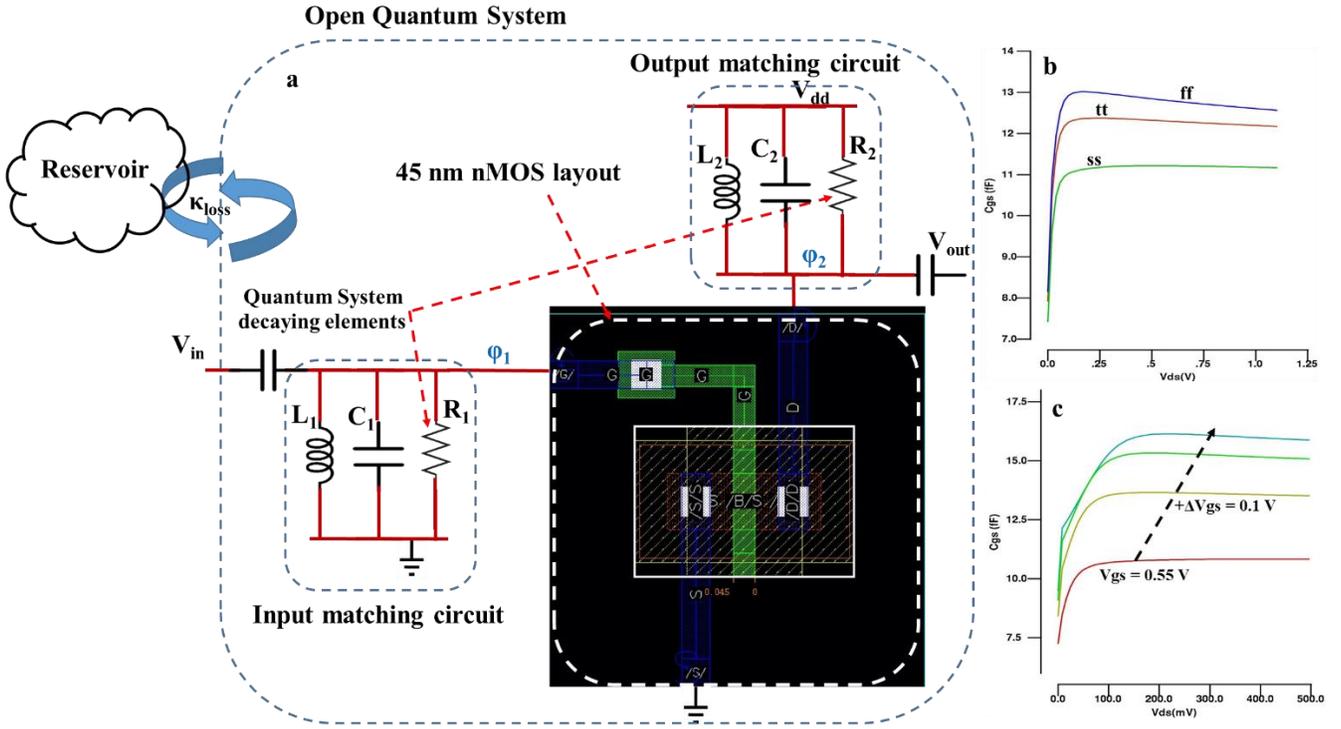

Fig. 1 Schematic of the defined system, a) two oscillator circuits coupled to 45 nm nMOS transistor (layout is depicted); b) gate-source capacitance changing at different operational corners versus $V_{ds}$ (V), $V_{gs}$ = 0.6 V; gate-source capacitance as a function of $V_{gs}$ (V) versus $V_{ds}$ (V), for all simulation in CAD area the operational temperature is 300 K.

## Quantum theory and dynamics equation of motion

Using full quantum theory, the circuit shown in Fig. 1 is analyzed. The data for the internal circuit of the nMOS transistor operating at deep cryogenic temperature is listed in Table. 1. The total Lagrangian of the circuit is theoretically derived, Eq. 1, in which the coordinate variables shown in Fig. 1, $\varphi_1$ and $\varphi_2$, are used.

$$L_c = \frac{C_{in}}{2}(V_{RF} - \dot{\varphi}_1)^2 - \frac{1}{2L_1}\varphi_1^2 + \overline{i_g^2}\varphi_1$$
$$+ \frac{C_{gs}}{2}\dot{\varphi}_1^2 + \frac{C_{gd}}{2}(\dot{\varphi}_1 - \dot{\varphi}_2)^2 - g_m\dot{\varphi}_1\varphi_2 \quad (1)$$
$$+ \left(\overline{i_{ds}^2} + \overline{i_d^2}\right)\varphi_2 + \frac{C_{ds}}{2}\dot{\varphi}_2^2 - \frac{1}{2L_2}\varphi_2^2$$

where $Q_1$ and $Q_2$ are the momentum conjugate variables derived using $Q_k = \partial L_s / \partial(\partial \varphi_k / \partial t)$ [29], k = 1,2; $g_m$ is the intrinsic transconductance of the transistor. The thermally generated noises by the resistors and the current source in the circuit are defined as $i_g^2 = 4K_BT/R_g$, $i_d^2 = 4K_BT/R_d$, and $i_{ds}^2 = 4K_BT\gamma g_m$, where $K_B$, T, and $\gamma$ respectively are the Boltzmann constant, operational temperature, and empirical constant. The classical Hamiltonian of the circuit is derived using the Legendre transformation [29, 36], which is expressed as:

$$H_c = \left\{\frac{1}{2C_1}Q_1^2 + \frac{1}{2L_1}\varphi_1^2 + \frac{1}{2C_2}Q_2^2 + \frac{1}{2L_2}\varphi_2^2\right\}$$
$$+ \left\{\frac{1}{2C_{12}}Q_1Q_2 + g_{12}Q_1\varphi_2 + g_{22}Q_2\varphi_2\right\} \quad (2)$$
$$+ \left\{V_{q1}Q_1 + V_{q2}Q_2 - \overline{i_{gs}^2}\varphi_1 - \left(G_1V_{RF} + \overline{i_{ds}^2} + \overline{i_d^2}\right)\varphi_2\right\}$$

In this equation, $C_1 = 0.5*(C_{p2}/C_M - 0.5/C_{1p})^{-1}$, $C_2 = 0.5*(C_{p1}/C_M - 0.5/C_{2p})^{-1}$, $C_{12} = 0.5*(2C_{gd}/C_M - 0.5/C_{12p})^{-1}$, $g_{12} = (g_mC_{gd}/C_M - g_{12p})$, $g_{22} = (g_mC_{p1}/C_M - g_{22p})$, $V_{q1} = V_{RF}(C_{in}C_{p2}/C_M - V_{q1p})$, $V_{q2} = V_{RF}(C_{in}C_{gd}/C_M - V_{q2p})$, where $C_{p1} = C_{in} + C_{gs} + C_{gd}$, $C_{p2} = C_{gd} + C_{ds}$, and $C_M^2 = C_{p2}C_{p1} - C_{gd}^2$. The defined variables as $C_{1p}$, $C_{2p}$, $C_{12p}$, $g_{12p}$, $g_{22p}$, $V_{q1p}$, $V_{q2p}$, and $G_1$ are defined in Appendix A, and the dc terms are ignored for simplicity.

In the following, one can derive the quantum Hamiltonian using the quantization procedure for the coordinates and the related momentum conjugates. The quadrature operators are defined as $Q_1 = -j(a_1 - a_1^+)(\hbar/2Z_1)^{0.5}$, $\varphi_1 = (a_1 + a_1^+)(\hbar Z_1/2)^{0.5}$ and $Q_2 = -j(a_2 - a_2^+)(\hbar/2Z_2)^{0.5}$, $\varphi_2 = (a_2 + a_2^+)(\hbar Z_2/2)^{0.5}$, where $(a_i, a_i^+)$ i = 1,2 are the first and second oscillator's annihilation and creation operators, respectively. The LC resonator impedances and the associated frequencies are defined, respectively, as $Z_1 = (L_g/C_1)^{0.5}$, $Z_2 = (L_d/C_2)^{0.5}$, and $\omega_1 = (L_gC_1)^{-0.5}$, $\omega_2 = (L_dC_2)^{-0.5}$. Finally, the quantum Hamiltonian in terms of the ladder operators is given by:

$$H_L = \left\{\hbar\omega_1\left(a_1^+a_1+\frac{1}{2}\right)+\hbar\omega_2\left(a_2^+a_2+\frac{1}{2}\right)\right\}$$
$$+\left\{-\frac{\hbar}{2}\frac{1}{C_{12}\sqrt{Z_1Z_2}}(a_1-a_1^+)(a_2-a_2^+)-\frac{j\hbar}{2}g_{12}\sqrt{\frac{Z_2}{Z_1}}(a_1-a_1^+)(a_2+a_2^+)-\frac{j\hbar}{2}g_{22}(a_2-a_2^+)(a_2+a_2^+)\right\} \quad (3)$$
$$+\left\{-jV_{q1}\sqrt{\frac{\hbar}{2Z_1}}(a_1-a_1^+)-jV_{q2}\sqrt{\frac{\hbar}{2Z_2}}(a_2-a_2^+)-\left[\overline{i_{ds}^2}+\overline{i_d^2}+G_1V_{RF}\right]\sqrt{\frac{\hbar Z_2}{2}}(a_2+a_2^+)-\overline{i_{gs}^2}\sqrt{\frac{\hbar Z_1}{2}}(a_1+a_1^+)\right\}$$

In this equation, three coefficients indicated by the red dashed squares are defined as $g_{q1q2} = -0.5/(C_{12}\sqrt{(Z_1Z_2)})$, $g_{q1p2} = -0.5g_{12}*\sqrt{(Z_2/Z_1)}$, and $g_{q2p2} = -0.5g_{22}$. These coefficients are the origin of the parametric amplifier. Indeed, the dynamics of parametric amplification are governed by these three critical factors that wield significant influence over the amplification process. Each of these factors (the first two of them have so critical effect) plays a pivotal role in determining the amplification characteristics and interactions within the system. The first key factor is denoted as $g_{q1q2}$, a coefficient that encapsulates the coupling between the charge operator of the first oscillator and the charge operator of the second oscillator. This factor is rooted in the capacitive coupling between the oscillators, and its manipulation introduces substantial changes to the overall system's quantum behavior. The second factor is $g_{q1p2}$, which represents the coupling strength between the charge operator of the first oscillator and the second oscillator flux operator. This factor stands out as the linchpin of the parametric amplification mechanism, as it intricately interacts with the system's intrinsic characteristics. Notably, $g_{q1p2}$ is further influenced by the intrinsic transconductance ($g_m$) of the system, as well as the coupling capacitors. The last important critical factor is the coupling coefficient between the charge operator of the second oscillator and its flux operator $g_{q2p2}$. This factor adds an additional dimension to the system's interplay, offering yet another degree of freedom for fine-tuning the amplification process.

Altering the discussed factors yields changes in the scattering matrix derived from the Heisenberg-Langevin equation. As a result, the intra-cavity modes undergo a transformation, consequently impacting the output and input modes of the system. This intricate chain of events culminates in the profound alteration of the system's gain, directly correlating with the manipulated factors. It's important to underscore that the interdependence of these factors introduces a nuanced level of control over the parametric amplification process. The intimate relationship between the factors and the system's quantum properties underscores the entwined nature of quantum mechanics and signal amplification. Through the delicate manipulation of the factors, this system offers a compelling platform for designing finely tuned parametric amplifiers. Indeed, the intricate interplay of the factors extends its influence beyond the realm of parametric amplification, directly shaping the quantum properties of the system. This connection between the control of amplification and the manipulation of quantum properties underscores the deeply intertwined relationship between quantum mechanics and signal processing. This point will be discussed in the results section.

To calculate the quantum properties of the system such as quantum discord, classical discord, Symplectic eigenvalue, and quantum purity, the dynamics of such an open quantum system using the master equation can be governed [29, 36]. Therefore, it is necessary to describe the state of the open quantum system in terms of the ensemble average state. Using the density matrix formalism, the related quantum state's probability distribution is defined. The complete form of the master equation is the Lindblad master equation defined as [29, 32]:

$$\dot{\rho}(t) = \frac{1}{j\hbar}[H,\rho(t)]$$
$$+\frac{1}{2}\sum_n\left[2C_n\rho(t)C_n^+ - \rho(t)C_n^+C_n - C_n^+C_n\rho(t)\right] \quad (4)$$

where $C_n$ is the collapse operator by which the quantum system is coupled to the environment, and H is the total Hamiltonian defined as $H_L + H_{env} + H_{int}$, where $H_{env}$ and $H_{int}$ are the environment and system-environment interaction Hamiltonian, respectively [32]. To derive Eq. 4, some approximations have been made [32], including 1. There is no correlation between the quantum system and environment at the initiating time; 2. The state of the environment doesn't significantly change since interacting with the quantum system; 3. The system and environment remain separable during the evolution; 4. The time scale of the environment is much shorter than the quantum system dynamics. By solving the Lindblad master equation expressed in Eq. 4, it will be shown how and by what degree, the parametric amplifier coefficients can affect the quantum properties of the system.

The quantum discord as a main quantifier is used to show the quantum correlation created by the system. The compact form of the quantum discord, $D(\rho_{AB})$, and classical discord $C(\rho_{AB})$ are defined as follows: $D(\rho_{AB}) = h(b) - h(v_-) - h(v_+) + h(\tau + \eta)$ and $C(\rho_{AB}) = h(a) - h(v_-) - h(v_+)$, where $v_\pm$ is the Symplectic eigenvalue of the covariance matrix (CM), and the function "h" is defined as $h(x) = (x+0.5)\log_2(x+0.5) - (x-0.5)\log_2(x-0.5)$ [33-35]. In addition, $b$ stands for the second oscillator output average photon number, and other quantities are defined as follows $\tau = d_{o12}^2/(b^2-1)$, and $\eta = a - (b \times d_{o12}^2/(b^2-1))$, where $a$ is the first oscillator output average photon number, and $d_{o12}$ is the phase-sensitive cross-correlation between two coupled oscillators. The last term in the quantum discord equation, $h(\tau + \eta)$, is the effect of the classical correlation depending on the type of measurement applied on the second oscillator [33-35].

In the following, we want to theoretically calculate the scattering matrix, using which the gain of the quantum system discussed can be calculated. To do so, it is necessary to apply the quantum Langevin equation (QLE) [29, 36], and calculate the time evolution of the intra-cavity modes ($a_1$ and $a_2$), which is as follows:

$$\dot{a}_1 = \left(-j\Omega_1 - \frac{\kappa_1}{2}\right)a_1 + \left(-jg_{q1q2} - g_{q1\varphi2}\right)a_2^+ - \gamma_{q1} - j\gamma_{\varphi 1} + a_{in-1}$$

$$\dot{a}_1^+ = \left(j\Omega_1 - \frac{\kappa_1}{2}\right)a_1^+ + \left(jg_{q1q2} - g_{q1\varphi2}\right)a_2 - \gamma_{q1} + j\gamma_{\varphi 1} + a_{in-1}$$

$$\dot{a}_2 = \left(-j\Omega_2 - \frac{\kappa_2}{2}\right)a_2 + \left(-jg_{q1q2} - g_{q1\varphi2}\right)a_1^+ - 2\gamma_{q2\varphi2}a_2^+ - \gamma_{q2} - j\gamma_{\varphi 2} + a_{in-2}$$

$$\dot{a}_2^+ = \left(j\Omega_2 - \frac{\kappa_2}{2}\right)a_2^+ + \left(jg_{q1q2} - g_{q1\varphi2}\right)a_1 - 2g_{q2\varphi2}a_2 - \gamma_{q2} + j\gamma_{\varphi 2} + a_{in-2}$$

(5)

where $\Omega_i = \omega_i - \omega$, $\kappa_i$, and $a_{in-i}$ (i = 1,2), are the detuning frequency, decay rate, and input noise, respectively. In this equation, some definitions are used as follows: $\gamma_{q1} = -V_{q1}\sqrt{(1/2\hbar Z_1)}$, $\gamma_{q2} = -V_{q2}\sqrt{(1/2\hbar Z_2)}$, $\gamma_{\varphi 2} = -(G_1 V_{RF} + i_{sc}^2 + i_d^2)\sqrt{(Z_2/2\hbar)}$. With transferring the equations into the Fourier domain, the intra-cavities mode relationship with the input of the system becomes:

$$\begin{bmatrix} j(\Omega_1+\omega)+\frac{\kappa_1}{2} & 0 & 0 & jg_{q1q2}+g_{q1\varphi2} \\ 0 & j(\omega-\Omega_1)+\frac{\kappa_1}{2} & -jg_{q1q2}+g_{q1\varphi2} & 0 \\ 0 & jg_{q1q2}+g_{q1\varphi2} & j(\Omega_2+\omega)+\frac{\kappa_2}{2} & 2g_{q2\varphi2} \\ -jg_{q1q2}+g_{q1\varphi2} & 0 & 2g_{q2\varphi2} & j(\omega-\Omega_2)+\frac{\kappa_2}{2} \end{bmatrix} \begin{bmatrix} a_1 \\ a_1^+ \\ a_2 \\ a_2^+ \end{bmatrix} = \begin{bmatrix} a_{in-1} \\ a_{in-1}^+ \\ a_{in-2} \\ a_{in-2}^+ \end{bmatrix}$$

(6)

The matrix represented in Eq.6 is the scattering matrix [$A_{sct}$], which is the confluence point of the intra-cavity modes and the input modes. In the scattering matrix, the coefficients indicated with the red dashed squares are the ones introduced in the previous section that affect critically the parametric amplifications. Indeed, those are the coefficients issued due to the nMOS transistor circuitry effect. In the following, using the input-output formula, $a_{out} = a_{in} + \sqrt{\kappa}*a$, one can define $a_{out}$ in terms of $a_{in}$ ([$a_{out}$] = (I + [$A_{sct}$]$^{-1}\sqrt{\kappa}$) *[$a_{in}$]), and calculate the gain of the quantum system. In the next step, the effect of the discussed coefficients on the quantum properties and also on the gain of the quantum system will be studied.

Table. 1 data used to model the internal circuit of nMOS transistor at the cryogenic temperature [27].

| Width (μm) | 42 |
|---|---|
| $C_{pg}$ (fF) | 13.65 |
| $C_{pd}$ (fF) | 12.11 |
| $L_g$ (pH) | 32.13 |
| $L_d$ (pH) | 32.24 |
| $L_s$ (pH) | 45.22 |
| $R_g$ (Ω) | 25.78 |
| $R_d$ (Ω) | 2.62 |
| $R_s$ (Ω) | 0.36 |

**Results and Discussions**

Parametric amplification is a process where an input signal at a specific frequency is amplified by exploiting the nonlinearity of a system. In the present coupled oscillator system, the interaction between the oscillators and the transistor's internal circuitry introduces the same effects. When the system is operating as a parametric amplifier, it essentially takes advantage of the energy exchange between different degrees of freedom within the system. In this case, the two oscillators are coupled through the transistor, and the energy exchange between the oscillators can be enhanced when specific conditions are met. This study has applied a comprehensive approach to analyzing a typical quantum electronic circuit, with a specific potential to be operated as a parametric amplifier. The gain of the quantum circuit becomes calculated by theoretical derivation of the system Hamiltonian and Heisenberg-Langevin equation to evaluate the system's dynamics, and finally using the input-output formula.

Analysis of Eq. 6 reveals that the gain of the system is closely tied to the intra-cavities photon numbers of the coupled oscillators. These photon numbers have a critical influence on the amplification process and are indicative of the energy present in the oscillators. The normalized oscillators' gain versus normalized frequency is simulated at the deep cryogenic temperature around 400 mK, and depicted in Fig. 2. The gain profile in Fig. 2 exhibits distinct peaks. These peaks align with the resonance frequencies of the associated oscillators, indicating that the amplification process is most efficient when the frequencies are matched. The study shows that the parametric amplifier coefficients are critical. The coefficients governing the interaction between the oscillators (charge and flux operators) significantly affect the gain profile (Fig. 2b, Fig.2c, and Fig.2d). By adjusting these coefficients, one can modulate the gain and potentially control the energy transfer between the oscillators. Through coefficient manipulation, it is found that it is possible to modify the amplitudes of the resonance peaks. This alteration of amplitudes can be likened to energy flowing back and forth between the oscillators, highlighting the intricate interplay of energy transfer and amplification.

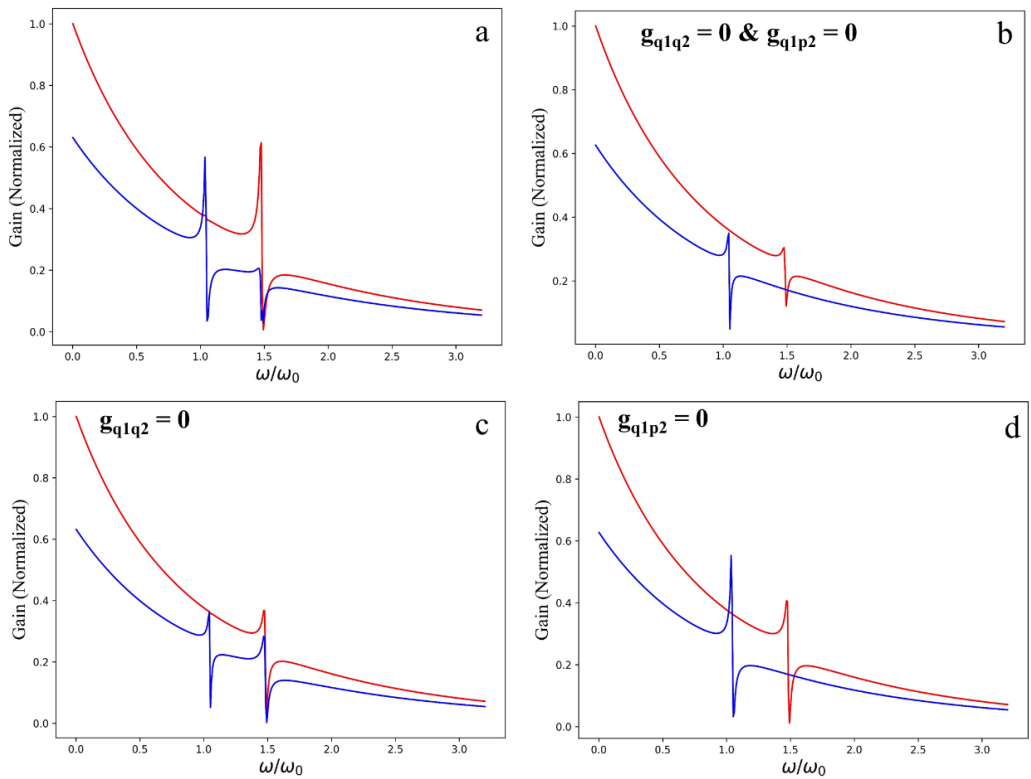

Fig. 2 The normalized first (blue) and second (red) oscillator gain vs. $\omega/\omega_0$; Tem = 300 mK, $\omega_0 = 2\pi*5.2$ GHz.

The overall analysis of Fig. 2 provides a thorough understanding of how the designed quantum electronic circuit functions as a potential parametric amplifier. The circuit is also simulated using CAD software, with the results visually presented in Fig. 3. It is done not only to verify the results derived using the quantum analysis but also to compare the results with the CAD simulations. To ensure the circuit's proper functionality, key parameters including $g_m/i_{ds}$, $V_{th}$ (transistor threshold voltage), and $g_m$ (transistor transconductance) are thoroughly examined and are graphically depicted in Fig. 3a. These analyses yielded results within expected norms. However, Fig. 3b showcases the circuit's gain and current flow simulation outcomes. An intriguing observation arises: alterations in gain, occurring at two distinct frequencies, coincide with concurrent changes in current flow (shown by the dashed rectangle in the figure). This phenomenon could be attributed to variations in impedance at these critical frequencies, suggesting a complex interplay between the circuit's amplification characteristics and its impedance profile. From the classical point of view, the change in gain at two distinct frequencies indicates resonance phenomena within the circuit. Resonance occurs when the circuit's natural frequency matches the frequency of the input signal. At resonance frequency, energy transfer is maximized, resulting in higher gain. When one tunes the circuit for increased gain at specific frequencies, it can lead to resonance and, consequently, changes in current flow due to enhanced energy transfer.

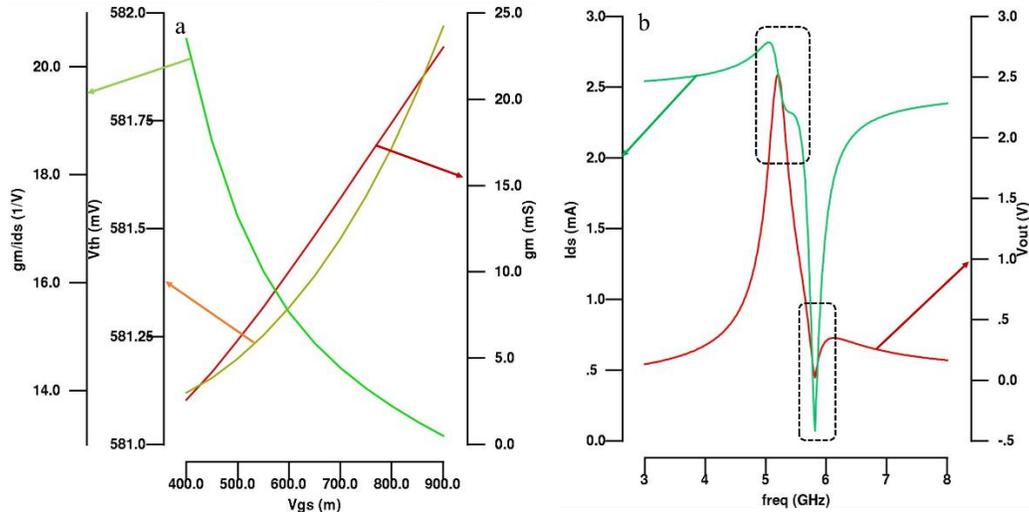

Fig. 3 DC and AC characterization of the circuit; a) $g_m/i_{ds}$ (1/V), $V_{th}$(V), and $g_m$(mA/V) vs. $V_{gs}$ (V); b) $i_{ds}$ (mA) and $V_{out}$ (V) vs. freq (GHz), Tem = 300 K.

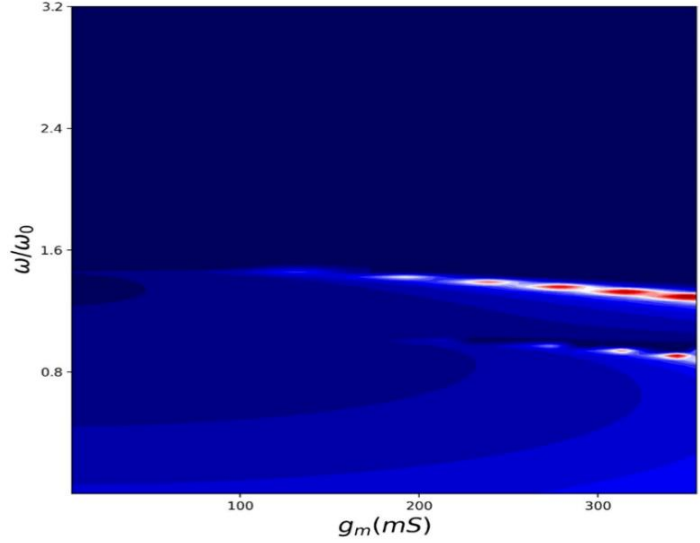

Fig. 4 the 2D normalized parametric amplifier gain vs. $g_m$ (mA/V) and $\omega/\omega_0$; Tem = 300 mK.

The 2D normalized parametric amplifier gain of the circuit is shown in Fig. 4. In the same way, it shows two distinct peaks, and their amplitude is increased with $g_m$ increasing. From the classical point of view, the gain of the simple circuit illustrated in Fig. 1, equals to $-g_m/g_{ds}$, where $g_{ds}$ is the drain-source conductance. Indeed, this figure shows the consistency between quantum theory and classical analysis. It is observed that increasing $g_m$ leads to higher gain aligns with the concept of amplification through nonlinearity. The gain in a parametric amplifier arises due to the system's ability to exchange energy between different modes with varying amplitudes. By increasing $g_m$, the parametric amplifier coefficients, $g_{q1q2}$, $g_{q1p2}$, and $g_{q2qp}$ are changed, which effectively enhances the nonlinearity of the system, making it more efficient in transferring energy between the oscillators.

In the following, as an interesting and complementary task, the effect of the parametric amplification coefficients is studied on the time evolution of the quantum properties such as quantum discord, classical discord, the smallest Symplectic eigenvalue, and quantum purity.

Quantum discord measures non-classical correlations that go beyond classical correlations based on shared information. It quantifies the quantum features of a system, including entanglement. From the comparison between Fig. 5a and Fig. 5b, it is observed that, at lower $g_m$, there's a significant difference between quantum discord $D(\rho_{AB})$ and classical correlations $C(\rho_{AB})$; for instance, where the quantum discord is maximized, the classical discord is minimized, and vice versa. The point indicates that the system exhibits non-classical quantum correlations. The reason may relate to the behavior of the smallest Symplectic eigenvalue ($d_-$) [34], which is illustrated in Fig. 5c. The behavior of the mentioned eigenvalue is a key factor in understanding the system's quantum properties; Gaussian state with covariance matrix is entangled if and only if $d_-<1$ [34]. The concept of entanglement is a cornerstone of quantum mechanics. Entanglement is a resource that is being explored for various quantum information tasks, such as quantum communication and quantum cryptography. So, $d_-$ decreasing at lower $g_m$, alongside the increase in quantum discord, suggests that these two quantities are correlated. The observation of this phenomenon reinforces the understanding that the system is transitioning from a classically correlated state to one that has non-local correlations between its parts.

At higher $g_m$ in contrast with lower $g_m$, the quantum discord and classical correlations behave approximately in the same way; it may indicate a transition or a change in the nature of correlations within the system. It is obvious since increasing $g_m$ leads to more trapped photons in the cavity, so it converts the quantum system to a classical one. Additionally, the reason may be attributed to the effect of the last term in the quantum discord formula, $h(\tau + \eta)$ [33], which is the effect of the classical correlation depending on the type of measurement applied on the second oscillator.

However, the most interesting phenomenon that this study found is indicated by the white dashed circles in Fig. 5a. It is shown that by increasing $g_m$, the width of frequencies (one can simply consider as bandwidth) in which the modes show non-classical correlation, is decreased. To explore this point, let's flashback to Fig. 1b and Fig. 1c, which are the $C_{gs}$ changing as the function of $V_{gs}$, $V_{ds}$, and operation speed of nMOS, and can be a perfect reason for the discussed point. It is shown in Fig. 1c that with increasing $V_{gs}$, meaning the increase of $g_m$, the slope of the change of the $C_{gs}$ is dramatically increased. Therefore, the change of $C_{gs}$ directly affects $C_{p1}$ and $C_M$ (Eq. 2), by which the parametric amplifier coefficients become influenced. For instance, larger capacitance arisen due to the tiny change in $V_{ds}$ in the gate-source leads to the dramatic decrease in $g_{q1q2}$ and $g_{q1p2}$. This is an interesting common ground that this study finds between the parametric amplifier and quantum properties of the system. Consequently, the manipulation of key factors serves as a powerful tool for shaping both the amplification potential and the quantum attributes of the system. This interplay opens avenues for tailoring quantum-enhanced signal processing, where the intricate dynamics of parametric amplification and quantum properties harmonize to pave the way for innovative applications in quantum information processing and quantum communication.

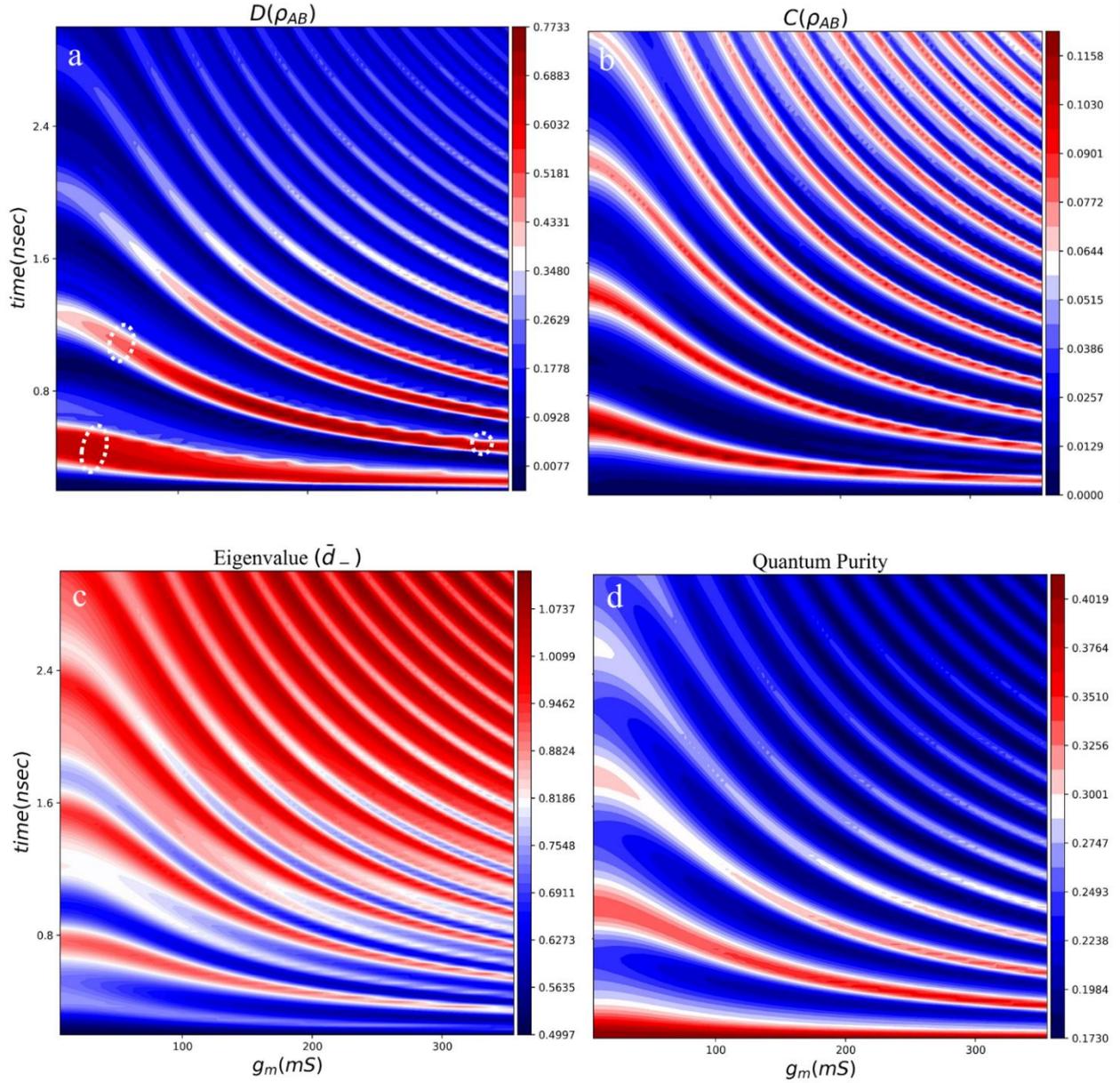

Fig. 5 The effect of the parametric amplification coefficients on system's quantum properties; a) quantum discord vs. time (nsec) and $g_m$ (mA/V); b) classical discord vs. time (nsec) and $g_m$ (mA/V); c) Symplectic eigenvalue vs. time (nsec) and $g_m$ (mA/V); d) quantum purity vs. time (nsec) and $g_m$ (mA/V); Tem = 300 mK.

Finally, the consistent behavior among quantum discord and quantum purity, shown in Fig. 5d, is a reflection of the underlying quantum nature of the system. It is interesting to note that an increase in quantum discord corresponds to an increase in quantum purity. This consistency underlines the holistic nature of quantum correlations and how they influence various aspects of a quantum system's behavior. As a pioneering quantum device, the designed circuit serves a dual purpose: not only does it amplify quantum signals as a quantum parametric amplifier, but it also enhances the inherent quantum properties of the signals; such as enhancing the non-classicality. In contrast to classical amplifiers, which merely amplify signals, this designed circuit functions as a quantum device. By carefully balancing the design parameters, one can create an effective parametric amplifier that simultaneously improves the quantum characteristics of the signals.

**Conclusions**
In a pioneering convergence of quantum theory, semiconductor technology, and cryogenic conditions, this study has illuminated a transformative path toward enhancing quantum properties and signal processing. By ingeniously designing a quantum electronic circuit within the confines of 45 nm CMOS technology, the potential to amplify quantum features at deep cryogenic temperatures has been harnessed. The manipulation of critical coefficients governing parametric amplification has unveiled not only a mechanism for signal enhancement but also a conduit for shaping quantum properties. This novel approach holds the promise of streamlining quantum signal processing, bridging technological gaps, and propelling quantum technology into unprecedented realms. We theoretically derived the system Hamiltonian and delved into the Heisenberg-Langevin equation to uncover the underlying dynamics. Through careful analysis, we identified specific coefficients that enable the system to function as a parametric amplifier. The calculated scattering matrix and subsequent gain simulations demonstrated that gain enhancement occurs at distinct frequencies, indicating the resonance behavior intrinsic to the system. This insight is crucial for optimizing the amplifier's performance for quantum signal amplification. Moreover, our study extended beyond gain enhancement to the exploration of quantum properties. By calculating quantum discord, classical discord, quantum purity, and the smallest Symplectic eigenvalue, it was revealed a web of correlations that illuminates the quantum behavior of the coupled oscillators. The consistent correlations among some of these quantities highlighted the interconnected nature of non-classical correlations and entanglement within the system.

Finally, by uncovering the relationships between interaction coefficients, gain enhancement, and quantum correlations, this study paves the way for more efficient and effective quantum devices that can amplify and process delicate quantum signals.

## References

1. M. Renger, S. Pogorzalek, Q. Chen, et al. Beyond the standard quantum limit for parametric amplification of broadband signals. npj Quantum Inf, 7, 160 (2021). https://doi.org/10.1038/s41534-021-00495-y.
2. E. Flurin, N. Roch, F. Mallet, M. H. Devoret, and B. Huard, Generating Entangled Microwave Radiation Over Two Transmission Lines, Phys. Rev. Lett. **109**, 183901 – Published 31 October 2012
3. Di Candia, R., Fedorov, K., Zhong, L. et al. Quantum teleportation of propagating quantum microwaves. EPJ Quantum Technol. 2, 25 (2015). https://doi.org/10.1140/epjqt/s40507-015-0038-9
4. B. Yurke, L. R. Corruccini, P. G. Kaminsky, L. W. Rupp, A. D. Smith, A. H. Silver, R. W. Simon, and E. A. Whittaker Observation of parametric amplification and deamplification in a Josephson parametric amplifier, Phys. Rev. A **39**, 2519 – Published 1 March 1989
5. F. Lecocq, L. Ranzani, G.A. Peterson, K. Cicak, A. Metelmann, S. Kotler, R.W. Simmonds, J.D. Teufel, and J. Aumentado, Microwave Measurement beyond the Quantum Limit with a Nonreciprocal Amplifier, Phys. Rev. Applied **13**, 044005 – Published 2 April 2020
6. C. Eichler, D. Bozyigit, C. Lang, M. Baur, L. Steffen, J. M. Fink, S. Filipp, and A. Wallraff , Observation of Two-Mode Squeezing in the Microwave Frequency Domain, Phys. Rev. Lett. **107**, 113601 – Published 6 September 2011
7. Samuel Boutin, David M. Toyli, Aditya V. Venkatramani, Andrew W. Eddins, Irfan Siddiqi, and Alexandre Blais, Effect of Higher-Order Nonlinearities on Amplification and Squeezing in Josephson Parametric Amplifiers, Phys. Rev. Applied **8**, 054030 – Published 15 November 2017
8. C. Eichler, Y. Salathe, J. Mlynek, S. Schmidt, and A. Wallraff, Quantum-Limited Amplification and Entanglement in Coupled Nonlinear Resonators, Phys. Rev. Lett. **113**, 110502 – Published 11 September 2014
9. C. Macklin, K. O'Brien, D. Hover, M. Schwartz, V. Bolkhovsky, X. Zhang, W. D. Oliver, I. Siddiqi, A near–quantum-limited Josephson traveling-wave parametric amplifier, Science, 2015, 350, 307-310, DOI: 10.1126/science.aaa852.
10. Luo, W., Cao, L., Shi, Y. *et al.* Recent progress in quantum photonic chips for quantum communication and internet. *Light Sci Appl* **12**, 175 (2023). https://doi.org/10.1038/s41377-023-01173-8
11. Di Candia, R., Minganti, F., Petrovnin, K.V. *et al.* Critical parametric quantum sensing. *npj Quantum Inf* **9**, 23 (2023). https://doi.org/10.1038/s41534-023-00690-z
12. M. Swathi and B. Rudra, "A Novel Approach for Asymmetric Quantum Error Correction With Syndrome Measurement," in *IEEE Access*, vol. 10, pp. 44669-44676, 2022, doi: 10.1109/ACCESS.2022.3170039.
13. L. Cochrane, T. Lundberg, D. J. Ibberson, L. A. Ibberson, L. Hutin, B. Bertrand, N. Stelmashenko, J. W. A. Robinson, M. Vinet, Ashwin A. Seshia, and M. F. Gonzalez-Zalba, Parametric Amplifiers Based on Quantum Dots, Phys. Rev. Lett. **128**, 197701 – Published 10 May 2022.
14. R. Yanagimoto, R. Nehra, R. Hamerly, E. Ng, A. Marandi, and H. Mabuchi, Quantum Nondemolition Measurements with Optical Parametric Amplifiers for Ultrafast Universal Quantum Information Processing, PRX Quantum **4**, 010333 – Published 29 March 2023.
15. E. Cha, N. Wadefalk, G. Moschetti, A. Pourkabirian, J. Stenarson and J. Grahn, "InP HEMTs for Sub-mW Cryogenic Low-Noise Amplifiers," in *IEEE Electron Device Letters*, vol. 41, no. 7, pp. 1005-1008, July 2020, doi: 10.1109/LED.2020.3000071.
16. E. Cha, N. Wadefalk, P. Nilsson, J. Schleeh, G. Moschetti, A. Pourkabirian, S. Tuzi, J. Grahn, "0.3–14 and 16–28 GHz Wide-Bandwidth Cryogenic MMIC Low-Noise Amplifiers," in *IEEE Transactions on Microwave Theory and Techniques*, vol. 66, no. 11, pp. 4860-4869, Nov. 2018, doi: 10.1109/TMTT.2018.2872566.
17. A Salmanogli, Entangled microwave photons generation using cryogenic low noise amplifier (transistor nonlinearity effects), Quantum Science and Technology 7, 045026, 2022.
18. A Salmanogli, Quantum correlation of microwave two-mode squeezed state generated by nonlinearity of InP HEMT Scientific Reports 13 (1), 11528, 2023.
19. A Salmanogli, Enhancing quantum correlation at zero-IF band by confining the thermally excited photons: InP hemt circuitry effect, Optical and Quantum Electronics 55 (8), 745, 2023.
20. A Salmanogli, Squeezed state generation using cryogenic InP HEMT nonlinearity, Journal of Semiconductors 44 (5), 052901, 2023.
21. A Salmanogli, HS Gecim, Accurate method to calculate noise figure in a low noise amplifier: Quantum theory analysis Microelectronics Journal 128, 105532, 2022.
22. A Salmanogli, Design of Ultra-Low Noise Amplifier for Quantum Applications (QLNA), arXiv preprint arXiv:2111.15358, 2021.
23. B. Prabowo, G. Zheng, M. Mehrpoo, B. Patra, P. Harvey-Collard, J. Dijkema, A. Sammak, G. Scappucci, E. Charbon, F. Sebastiano, L. M. K. Vandersypen, M. Babaie, A 6-to-8GHz 0.17mW/Qubit Cryo-CMOS Receiver for Multiple Spin Qubit Readout in 40nm CMOS Technology, ISSCC 2021 / SESSION 13 / CRYO-CMOS FOR QUANTUM COMPUTING / 13.3.

24. M. Mehrpoo, B. Patra, J. Gong, P. A. 't Hart, J. P. G. van Dijk, H. Homulle, G. Kiene, A. Vladimirescu, F. Sebastiano, E. Charbon, M. Babaie, Benefits and Challenges of Designing Cryogenic CMOS RF Circuits for Quantum Computers, 978-1-7281-0397-6/19/$31.00 ©2019 IEEE.

25. E. Charbon, Cryo-CMOS Electronics for Quantum Computing Applications, 978-1-7281-1539-9/19/$31.00 ©2019 IEEE.

26. Stefano Pellerano, Sushil Subramanian, Jong-Seok Park, Bishnu Patra, Todor Mladenov, Xiao Xue, Lieven M. K. Vandersypen, Masoud Babaie, Edoardo Charbon, Fabio Sebastiano, Cryogenic CMOS for Qubit Control and Readout, IEEE CICC 2022.

27. Yanfei Shen, Jie Cui, Saeed Mohammadi, An accurate model for predicting high frequency noise of nanoscale NMOS SOI transistors, Solid-State Electronics 131 (2017) 45–52.

28. L. Negre, D. Roy, S. Boret, P. Scheer, D. Gloria, Advanced 45nm MOSFET Small-signal equivalent circuit aging under DC and RF hot carrier stress, 978-1-4244-9111-7/11/$26.00 ©2011 IEEE, IRPS11-814 HV.1.4 Auth.

29. M. O. Scully, M. S. Zubairy, Quantum Optics, Cambridge University Press, UK, 1997.

30. A Salmanogli, Modification of a plasmonic nanoparticle lifetime by coupled quantum dots, Physical Review A 100 (1), 013817, 2019.

31. A Salmanogli, Quantum analysis of plasmonic coupling between quantum dots and nanoparticles, Phys. Rev. A 94, 2016.

32. J. R. Johansson, P. D. Nation, and F. Nori 2013 Comp. Phys. Comm. 184, 1234.

33. S.Pirandola, G. Spedalieri, S. L. Braunstein, N. J. Cerf, and S. Lloyd, Optimality of Gaussian Discord, Phys. Rev. Lett. 113, 140405-5, 2014.

34. G. Adesso and A. Datta, Quantum versus Classical Correlations in Gaussian States, Phys. Rev. Lett. 105, 030501-4, 2010.

35. P. Giorda and M. G. A. Paris, Gaussian Quantum Discord, Phys. Rev. Lett. 105, 020503-4, 2010.

36. B. Huttner, S. M. Barnett 1992 *Phys Rev A,* **46**, 4306.

## Appendix A:

In this appendix all of the variables used in the main article's equations listed as follows: $C_{1p}$ [F], $C_{2p}$ [F], $C_{12p}$ [F], $g_{12p}$ [1/s], $g_{22p}$ [1/s], $V_{q1p}$ [v], $V_{q2p}$ [v] and $G_1$ [A/V] are given by:

$$\frac{1}{2C_{1p}} = \frac{C_{p1}C_{p2}^2 - C_{gd}^2 C_{p2}}{2|C_M|^2}$$

$$\frac{1}{2C_{2p}} = \frac{C_{p1}^2 C_{p2} - C_{p1}C_{gd}^2}{2|C_M|^2}$$

$$\frac{1}{C_{12p}} = \frac{-C_{gd}(C_{p1}C_{p2} + C_{gd}^2) + 2C_{p1}C_{gd}C_{p2}}{|C_M|^2}$$

$$g_{12p} = \frac{-g_m C_{gd} C_{p1} C_{p2} + 2g_m C_{p1}^2 C_{p2}}{|C_M|^2} - \frac{g_m C_{p2}}{|C_M|}$$

$$g_{22p} = \frac{g_m C_{gd}^2 C_{p1} - g_m C_{gd}(C_{p1}C_{p2} + C_{gd}^2)}{|C_M|^2} - \frac{g_m C_{gd}}{|C_M|}$$

$$V_{q1p} = \frac{C_{in}C_{p1}C_{p2}^2 - C_{gd}^2 C_{in} C_{p2}}{|C_M|^2}$$

$$V_{q2p} = \frac{2C_{in}C_{p1}C_{p1}C_{gd} - C_{gd}(C_{gd}^2 C_{in} + C_{in}C_{p1}C_{p2})}{|C_M|^2}$$

$$G_1 = \frac{2g_m C_{in} C_{p1} C_{p1} C_{gd} - g_m C_{gd}(C_{gd}^2 C_{in} + C_{in}C_{p1}C_{p2})}{|C_M|^2} - \frac{g_m C_{in} C_{p2}}{|C_M|}$$